\documentclass[aps,amsmath,amssymb, nofootinbib,  twocolumn, 
superscriptaddress,prx]{revtex4-1}

\usepackage{graphicx}
\usepackage{color}
\usepackage{bbold}

\newcommand{\zb}{{\boldsymbol z}}

\newcommand{\bit}{\begin{itemize}}
\newcommand{\eit}{\end{itemize}}

\newcommand{\f}{\frac}
\renewcommand{\>}{\right\rangle}
\newcommand{\<}{\left\langle}
\newcommand{\ba}{\begin{align}}
\newcommand{\ea}{\end{align}}
\newcommand{\be}{\begin{equation}}
\newcommand{\ee}{\end{equation}}
\newcommand{\bi}{\begin{itemize}}
\newcommand{\ei}{\end{itemize}}
\newcommand{\lf}{\left(}
\newcommand{\ri}{\right)}
\newcommand{\dd}{\mathrm{d}}

\newcommand{\OO}{\mathcal{O}}

\newcommand{\Tr}{\operatorname{Tr}}

\newcommand{\cp}{\mathrm{CP}}
\newcommand{\nccp}{\mathrm{NCCP}}

\def\+{\dagger}

\newcommand{\nx}{\widetilde N_x}
\newcommand{\fx}{\widetilde \varphi_x}

\begin{document}

\newcommand{\bra}[1]{\< #1 \right|}
\newcommand{\ket}[1]{\left| #1 \>}

\title{Emergent $\mathrm{SO}(5)$ Symmetry at the N\'eel to Valence--Bond--Solid Transition}

\author{Adam Nahum} 
\affiliation{Department of Physics, Massachusetts Institute of Technology, Cambridge, MA 02139, USA}
\author{P. Serna}
\affiliation{Theoretical Physics, Oxford University, 1 Keble Road, Oxford OX1 3NP, United Kingdom}
\author{J. T. Chalker}
\affiliation{Theoretical Physics, Oxford University, 1 Keble Road, Oxford OX1 3NP, United Kingdom}
\author{M. Ortu\~no}
\author{A. M. Somoza}
\affiliation{Departamento de F\'isica -- CIOyN, Universidad de Murcia, Murcia 30.071, Spain}
\date{\today}

\begin{abstract}
\noindent
We show numerically that the `deconfined' quantum critical point between the N\'eel antiferromagnet and the columnar valence--bond--solid, for a square lattice of spin-1/2s, has an emergent $\mathrm{SO}(5)$ symmetry. This symmetry allows the N\'eel vector and the valence-bond-solid order parameter to be rotated into each other. It is a remarkable 2+1--dimensional analogue of the ${\mathrm{SO}(4)= [\mathrm{SU}(2)\times \mathrm{SU}(2)]/\mathbb{Z}_2}$ symmetry that appears in the scaling limit for the spin--1/2 Heisenberg chain. The emergent $\mathrm{SO}(5)$ is strong evidence that the phase transition in the 2+1D system is truly continuous, despite the violations of finite-size scaling observed previously in this problem. It also implies surprising relations between correlation functions at the transition. The symmetry enhancement is expected to apply generally to the critical two-component Abelian Higgs model (non-compact $\cp^1$ model). The result indicates that in three dimensions there is an $\mathrm{SO}(5)$-symmetric conformal field theory which has no relevant singlet operators, so is radically different to conventional Wilson-Fisher-type conformal field theories.
\end{abstract}

\maketitle

\noindent
Many condensed matter systems show higher symmetry in the infrared than they do in the ultraviolet. The liquid-gas critical point is a classical example: although there is no microscopic $\mathbb{Z}_2$ symmetry exchanging liquid-like and gas-like configurations, the fixed point has an emergent $\mathbb{Z}_2$ symmetry and is simply the Ising fixed point. Microscopically this fixed point is perturbed by operators which break the $\mathbb{Z}_2$ symmetry, but it nevertheless governs the critical behaviour because these perturbations are irrelevant under the renormalisation group.

To reach this critical point two variables, say temperature and the pressure, must be tuned. The spin-1/2  Heisenberg chain provides an example of emergent symmetry without such fine-tuning in a quantum setting. The ground state of this model is well known to be critical. Its microscopic symmetries are $\mathrm{SU}(2)$ spin rotations, together with spatial symmetries. However the scaling limit of the spin-1/2 chain is the $\mathrm{SU}(2)_1$ Wess-Zumino-Witten conformal field theory \cite{affleck cts symmetries}, and this has an $\mathrm{SO}(4) = [ \mathrm{SU}(2)\times \mathrm{SU}(2)]/\mathbb{Z}_2$ symmetry which is much larger than the global symmetry present microscopically. 

Physically, this arises as follows  \cite{tsvelik book}. The N\'eel vector $\vec N$ has three components. There is also a spin-Peierls order parameter $\varphi$ which distinguishes between the two different patterns of dimer (singlet) order, and which changes sign under appropriate reflections or translations. We may form the 4-component superspin ${\vec \Phi = (\vec N, \varphi)}$, and the emergent $\mathrm{SO}(4)$ corresponds to rotations of this vector. Although the dimer  and N\'eel order parameters are utterly inequivalent microscopically, a symmetry between them arises in the infra-red. Technically, this again relies on the $\mathrm{SO}(4)$--breaking perturbations of the conformal field theory being irrelevant or marginally irrelevant.

Naively one might expect this phenomenon to be special to one spatial dimension, where the enlarged symmetry is related to special properties of 2D conformal invariance (the doubling of conserved currents \cite{affleck cts symmetries}). We show here however that an analogous symmetry enhancement occurs for the spin-1/2 magnet on the square lattice, at the celebrated `deconfined' quantum critical point \cite{deconfined critical points, quantum criticality beyond, critically defined}.  This is a transition between the antiferromagnetically ordered N\'eel state and a columnar valence-bond-solid (VBS), and is reached by tuning a single parameter. For example, the nearest-neighbour Heisenberg model can be driven into the VBS using either a four-spin interaction \cite{Sandvik JQ} or a next--nearest--neighbour exchange \cite{wang gu et al dcp, sandvik FSS}. The emergent symmetry we put forward is an  $\mathrm{SO}(5)$ which mixes the components of the N\'eel vector $\vec N$, which has three components, and the VBS order parameter $\vec \varphi$, which has two. We test it by examining the joint probability distribution for these quantities.

Numerically, the critical behaviour can be  studied efficiently with a 3D classical loop model \cite{deconfined long paper}, and we use this approach here. The order of the transition has been controversial as a result of violations of conventional finite--size scaling \cite{Sandvik logs, Kawashima deconfined criticality, Jiang et al, deconfined criticality flow JQ, Kaul SU(3) SU(4), Banerjee et al} which we discussed in detail previously \cite{deconfined long paper}. We will return to this below, arguing that the present results  support the continuity of the transition.

The N\'eel--VBS transition is usually described with the non-compact $\cp^1$ ($\nccp^1$) Lagrangian \cite{Motrunich Vishwanath, quantum criticality beyond},
\be\label{NCCP1 lagrangian}
\mathcal{L} = |(\nabla - i A) \zb|^2 + \kappa (\nabla \times A)^2 + \mu |\zb|^2 + \lambda |\zb|^4.
\ee
The two-component bosonic spinon field $\zb = (z_1, z_2)$ is related to the N\'eel vector $\vec N$ by ${\vec N = \zb^\dag \vec \sigma \zb}$. The $\mathrm{U}(1)$ gauge field $A_\mu$ is related by duality to the VBS order parameter $\vec \varphi = (\varphi_x, \varphi_y)$ which distinguishes the different columnar singlet patterns \cite{levin senthil, quantum criticality beyond}.  Although we will use the language of the N\'eel--VBS transition, our conclusions apply more generally to the  above field theory, and indicate that it flows to an $\mathrm{SO}(5)$--symmetric fixed point at the critical value of $\mu$. In the language of this 3D gauge theory, the VBS order parameter is the operator $\mathcal{M} = \varphi_x + i \varphi_y$ which inserts a Dirac monopole in $A_\mu$ \cite{quantum criticality beyond, haldane 2+1,read sachdev prl}.

$\mathrm{SO}(5)$ symmetry cannot be made explicit in the formulation of Eq.~\ref{NCCP1 lagrangian}. Fortunately, Senthil and Fisher \cite{senthil fisher competing orders}, building on work of Tanaka and Hu \cite{tanaka hu}, have argued that an alternative field theory describes the N\'eel--VBS transition and is equivalent to Eq.~\ref{NCCP1 lagrangian}. This is a nonlinear $\sigma$-model (NL$\sigma$M) for the five-component superspin
\be
\vec \Phi = (N_x, N_y, N_z, \varphi_x, \varphi_y),
\ee
augmented with: (i) anisotropies that break the global symmetry from $\mathrm{SO}(5)$  down to the spin rotation and spatial symmetries present microscopically; and (ii)  a topological Wess--Zumino--Witten term at level one \cite{current algebra, abanov weigmann}, which is analogous to that in the CFT for the spin chain. The leading anisotropy plays the role of the mass term in Eq.~\ref{NCCP1 lagrangian}: it drives the transition between the N\'eel and VBS ordered phases. 

The NL$\sigma$M formulation makes the emergent symmetry a more natural possibility, since it could arise at the critical point if all the higher anisotropies happen to be renormalisation--group irrelevant. We will discuss below the phase diagram for the NL$\sigma$M (with WZW term) which is implied by this conjecture.

In previous work we characterised various observables at the deconfined transition  in detail, using a three-dimensional loop representation to reach system sizes up to $L=512$. See Ref.~\cite{deconfined long paper} for details of the model, which is in the N\'eel phase for coupling $J<J_c$ and the VBS phase for $J>J_c$, with $J_c = 0.088501(3)$.  We found a remarkable similarity between the critical  N\'eel and VBS correlation functions. The anomalous dimensions determined from the correlators  at separations $r\ll L$ are $\eta_\text{N\'eel}= { 0.259(6)}$ and $\eta_\text{VBS}= { 0.25(3)}$ \cite{eta footnote}; the two correlators also behave similarly in the regime $r\sim L$ (despite the scaling violations discussed in Ref.~\cite{deconfined long paper}). This suggests searching for an emergent $\mathrm{SO}(5)$ symmetry that would explain these apparent coincidences. 

\begin{figure}[t]
 \begin{center}
 \includegraphics[width=0.95\linewidth]{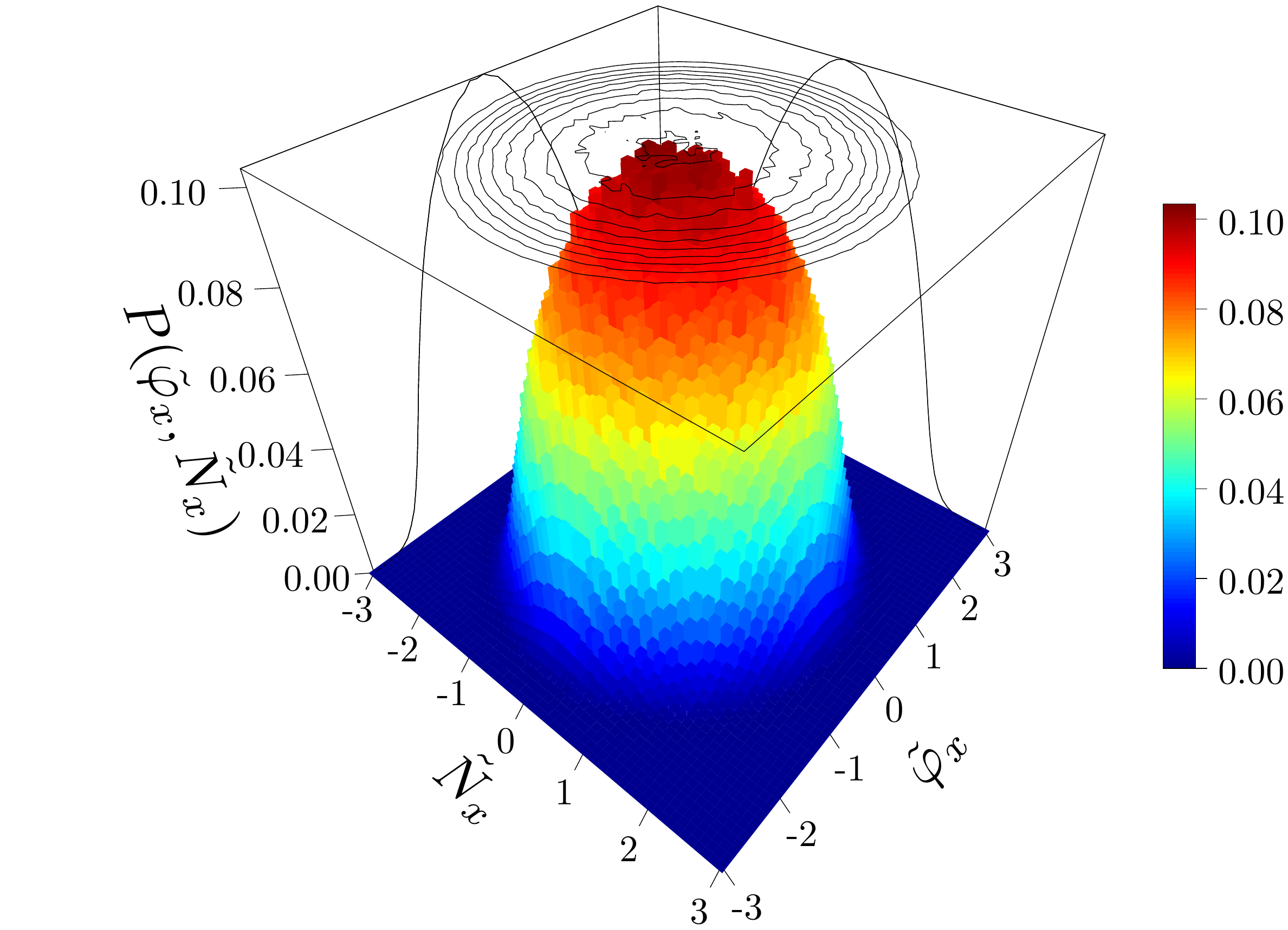}
 \end{center}
\caption{The joint probability distribution $P(\widetilde N_x, \widetilde\varphi_x)$, after rescaling $N_x$ and $\varphi_x$ to have unit variance, in a critical system of size $L=100$. Upper plane shows contour plot.}
 \label{probability distribution}
\end{figure}

\textit{Probability distribution.} Consider the joint distribution for the N\'eel and VBS order parameters in a system of linear size $L$. If $\mathrm{SO}(5)$ emerges, then this will be a function only of  {$\vec\Phi^2 = \vec N^2 + \vec \varphi^2$} at the critical point (after a trivial rescaling of $\vec\varphi$). Spin rotation symmetry of course already guarantees that the distribution depends on $\vec N$ only via $\vec N^2$. Also, while microscopic spatial symmetry only allows $\vec\varphi$ to rotated by multiples of $\pi/2$, it is well established numerically that symmetry under continuous $\mathrm{U}(1)$ rotations of $\vec \varphi$ emerges close to the transition \cite{lou sandvik kawashima, block melko kaul}. This was checked for the present model in Ref.~\cite{deconfined long paper}  (see also App.~\ref{additional moment data}, and see Ref.~\cite{Misguich Pasquier Alet} for related phenomena).  The crucial point is therefore whether the distribution is invariant under $\mathrm{U}(1)$ rotations that mix a component of $\vec\varphi$ with a component of $\vec N$.

\begin{figure}[b]
 \begin{center}
  \includegraphics[width=0.83\linewidth]{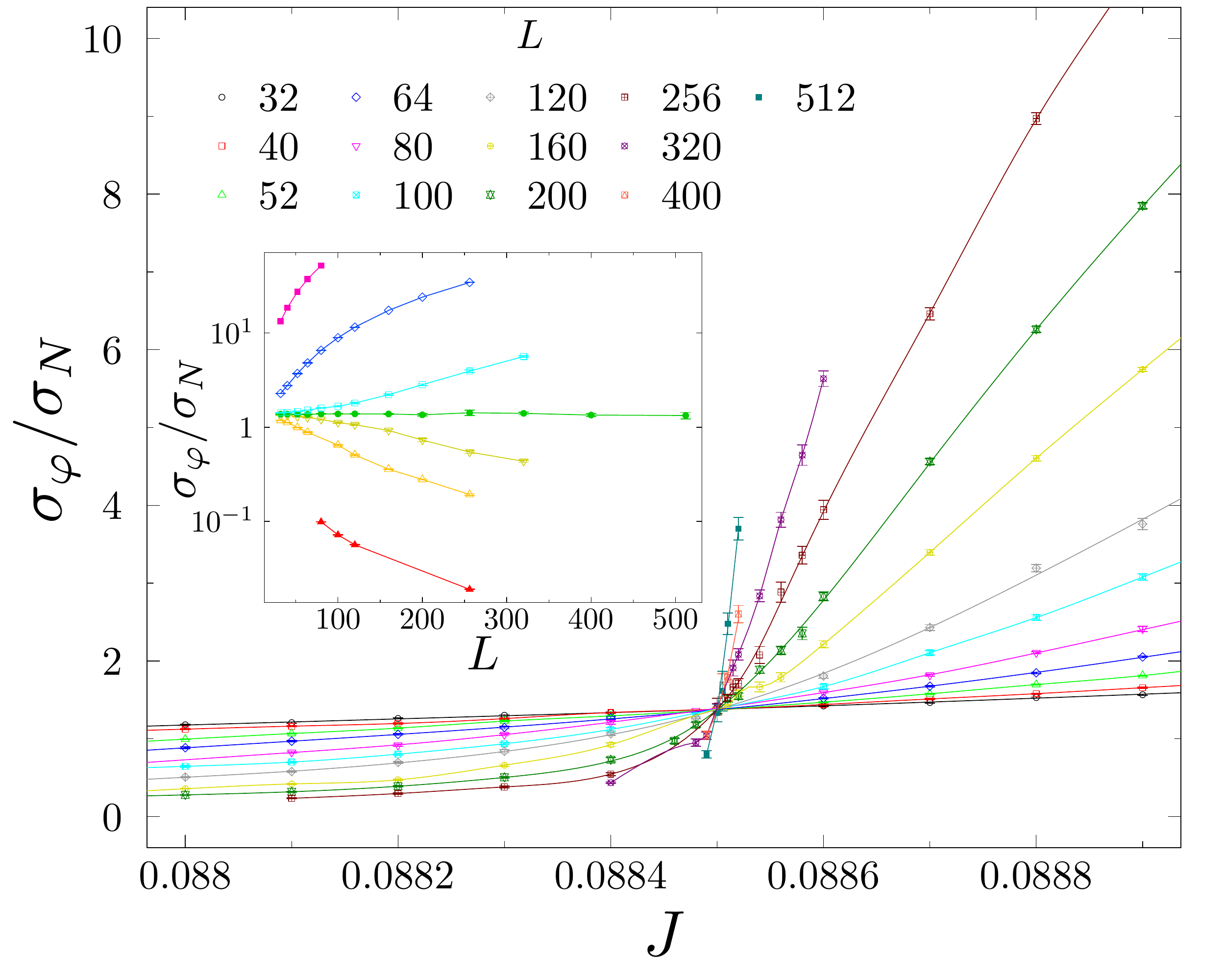}
 \end{center}
\caption{Main panel: variance ratio $\sigma_\varphi / \sigma_N$ plotted against $J$ for various $L$. Curves cross at $J_c$ as expected from $\mathrm{SO}(5)$ symmetry. Inset: same quantity as a function of $L$ for several $J$ around $J_c\simeq 0.0885$ (key in Fig.~\ref{moments as a function of L}).}
 \label{variance ratio}
\end{figure}

Let the standard deviations of $N_x$ and $\varphi_x$ be denoted $\sigma_N$ and $\sigma_\varphi$ respectively, and use a tilde to denote quantities rescaled to have unit variance: ${\widetilde N_x  = N_x / \sigma_N}$ and ${\widetilde \varphi_x = \varphi_x / \sigma_\varphi}$. Fig.~\ref{probability distribution} shows the joint probability distribution for these quantities at the critical point $J=J_c$ in a system of size $L=100$. The visual evidence for emergent symmetry between N\'eel and VBS is striking.

Before turning to a quantitative analysis of the distribution, a  basic test is that the variances $\sigma_N$ and $\sigma_\varphi$ of the order parameters depend on system size in the same way at criticality \cite{scaling dim footnote}, i.e. that $\sigma_\varphi/\sigma_N$ is $L$-independent at $J_c$.  In Fig.~\ref{variance ratio} this is confirmed to high precision over a wide range of lengthscales.  Plots of $\sigma_\varphi/\sigma_N$ versus $J$ for different $L$-values cross at $J_c$ \cite{scaling collapse footnote} (the value of $\sigma_\varphi/\sigma_N$ at $J_c$ depends on the nonuniversal normalisation of the lattice operators).

\begin{figure}[b]
 \begin{center}
  \includegraphics[width=0.83\linewidth]{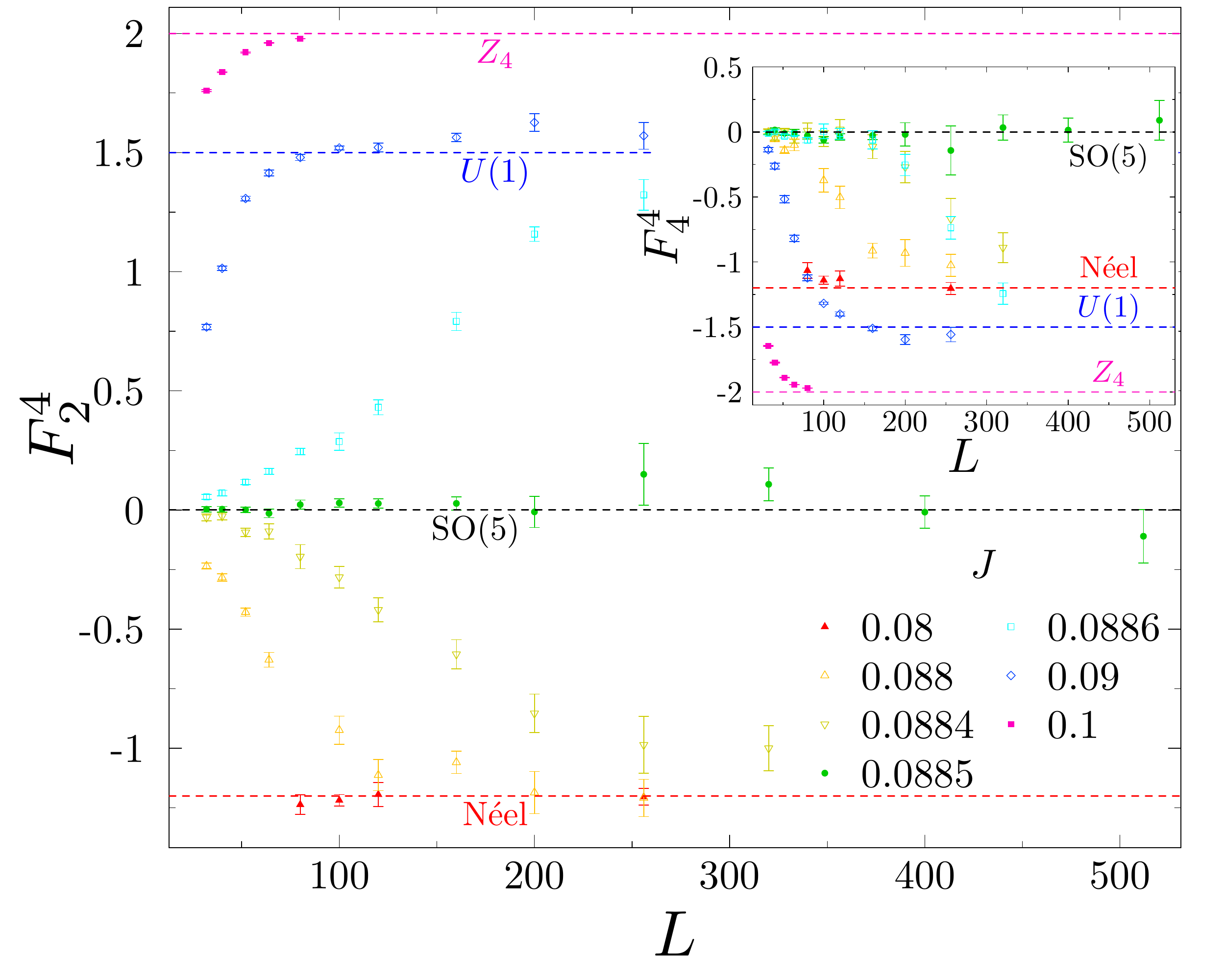}
 \end{center}
\caption{Moments $F_2^4$ (main panel) and $F_4^4$ (inset) defined in Eq.~\ref{two harmonics}, shown as a function of $L$ for a few values of $J$ at and close to $J_c\simeq 0.0885$. Dashed lines indicate expected values in the N\'eel phase, at the $\mathrm{SO}(5)$--symmetric critical point, and in the $\mathrm{U}(1)$ and $\mathbb{Z}_4$--symmetric regimes of the VBS phase.}
 \label{moments as a function of L}
\end{figure}

For a quantitative analysis of the probability distribution we examine the moments
\be\label{moments eq}
F_{\ell}^a =\langle r^a \cos \lf  \ell \theta \ri \rangle,
\ee
 where $(\widetilde N_x, \widetilde \varphi_x) = r \, (\cos\theta, \sin\theta)$. Emergent symmetry requires  these to vanish for $\ell>0$. We have computed $F_2^4$ and $F_4^4$ for large sizes:
  \ba\label{two harmonics}
 F_2^4 &= \< \nx^4 - \fx^4 \>,
&
F^4_4&= \<\nx^4-6\nx^2 \fx^2 + \fx^4 \>.
 \end{align}
Fig.~\ref{moments as a function of L} shows these as a function of $L$ at and close to the critical point. Both are consistent with zero for ${J=J_c}$ over the whole range of $L$. The expected values in the adjacent phases (including the regime of weak VBS order, where there is an effective $\mathrm{U}(1)$ symmetry for rotations of $\vec\varphi$) are also indicated in the figure (details in App.~\ref{additional moment data}).

\begin{figure}[t]
 \begin{center}
  \includegraphics[width=0.83\linewidth]{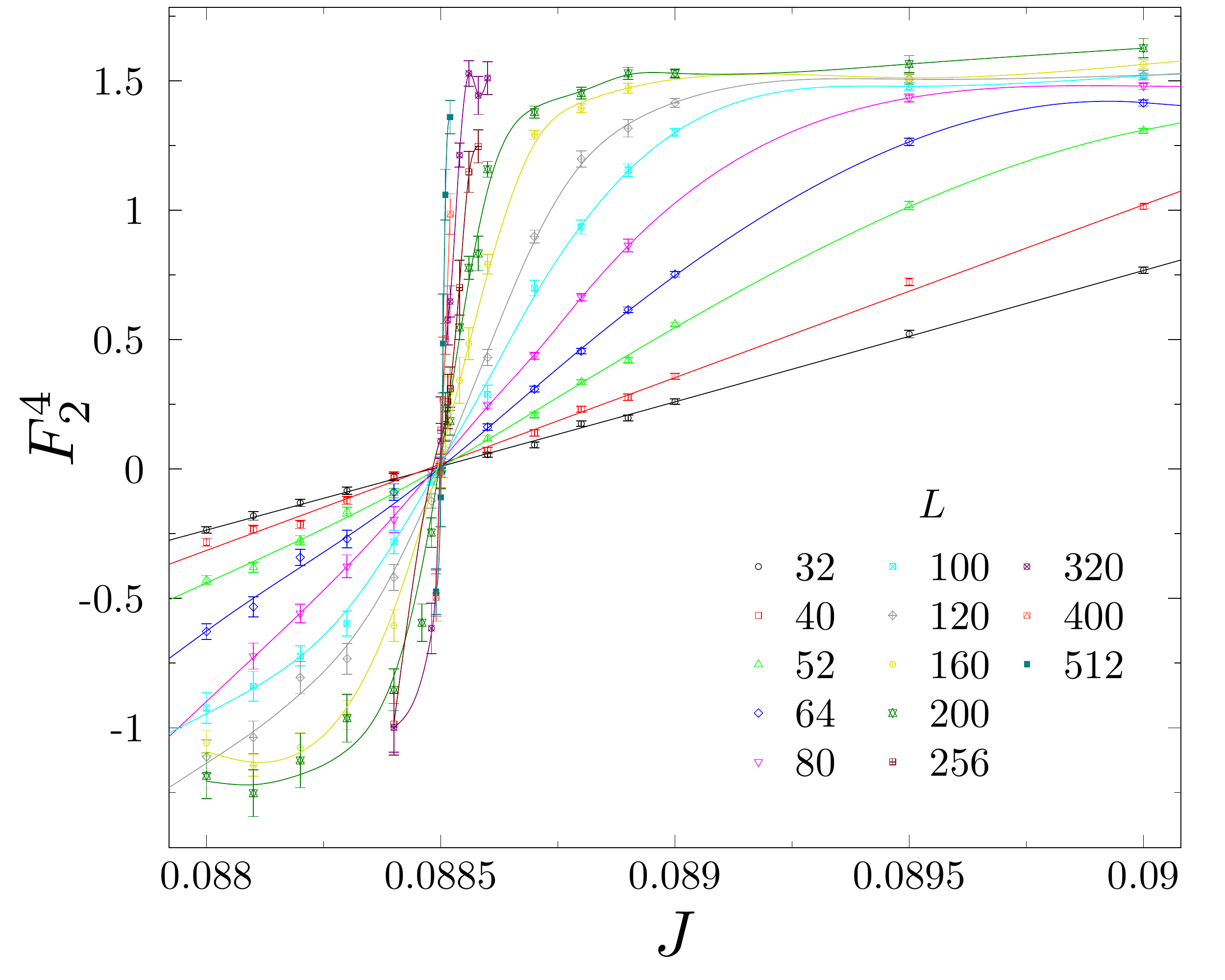}
 \end{center}
\caption{$F_2^4$ (Eq.~\ref{two harmonics}) as a function of $J$ for various sizes, showing a crossing at $F_2^4=0$, $J=J_c$ as expected for $\mathrm{SO}(5)$ symmetry.}
 \label{moment as a function of J}
\end{figure}

In Fig.~\ref{moment as a function of J} we show $F_2^4$ as a function of $J$: note the very clearly defined crossing at $J=J_c$, $F_2^4= 0$. Further moments are shown  for $L=100$ in App.~\ref{additional moment data}. It should be noted that the critical distribution is markedly non-Gaussian, with nonvanishing higher cumulants such as  ${[\langle N_x^4\rangle-3\langle N_x^2\rangle^2]/\langle N_x^2\rangle^2 = -0.7549(13)}$ (for $L=100$).

\textit{Equalities between scaling dimensions.} In addition to the equivalence between N\'eel and VBS vectors (manifested in the joint distribution and the anomalous dimensions),  $\mathrm{SO}(5)$ has consequences for operators transforming in higher representations. Take the leading operators in the symmetric two-- and four--index representations:
\begin{align}\notag
\OO_{ab}^{(2)} &  = \Phi_a \Phi_b  - \tfrac{1}{5} \delta_{ab} \Phi^2, &
  \OO_{abcd} ^{(4)}& =\Phi_a \Phi_b \Phi_c \Phi_d 
- C_{abcd}.
\end{align}
The subtractions \cite{trace term} ensure the operators transform irreducibly. $\OO^{(2)}$ is relevant, with scaling dimension $x_2 < 3$. In fact a component of $\OO^{(2)}$ is the operator $\OO_J$ which drives us through the N\'eel--VBS transition as we vary $J$, by favouring one or the other order ($\OO_J$ therefore plays the role of the mass term in Eq.~\ref{NCCP1 lagrangian}):
\be\label{relevant anistropy}
\OO_J = \tfrac{5}{2}\, {\sum}_{a=1}^3  \OO^{(2)}_{aa} = \vec N^2 - \tfrac{3}{2}\vec\varphi^2.
\ee
Remarkably, various \textit{a priori} unrelated operators share the same scaling dimension $x_2$ since they are also components of $\OO^{(2)}$. These include the spin--quadrupole moments, {$N_a N_b-\delta_{ab}\vec N^2/3$}, and the relevant \cite{block melko kaul} operators {$\varphi_a \varphi_b-\delta_{ab}\vec\varphi^2/2$} which in the $\nccp^{1}$ language insert `strength-two' monopoles \cite{rectangular anisotropy}. Even more oddly, the same scaling dimension controls $\varphi_a N_b$, despite the fact that microscopically this operator is as dissimilar from $\OO_J$ as possible --- $\varphi_a N_b$ transforms under both spin and spatial symmetries, while $\OO_J$ is invariant under them.

\begin{figure}[b]
 \begin{center}
 \includegraphics[width=0.75\linewidth]{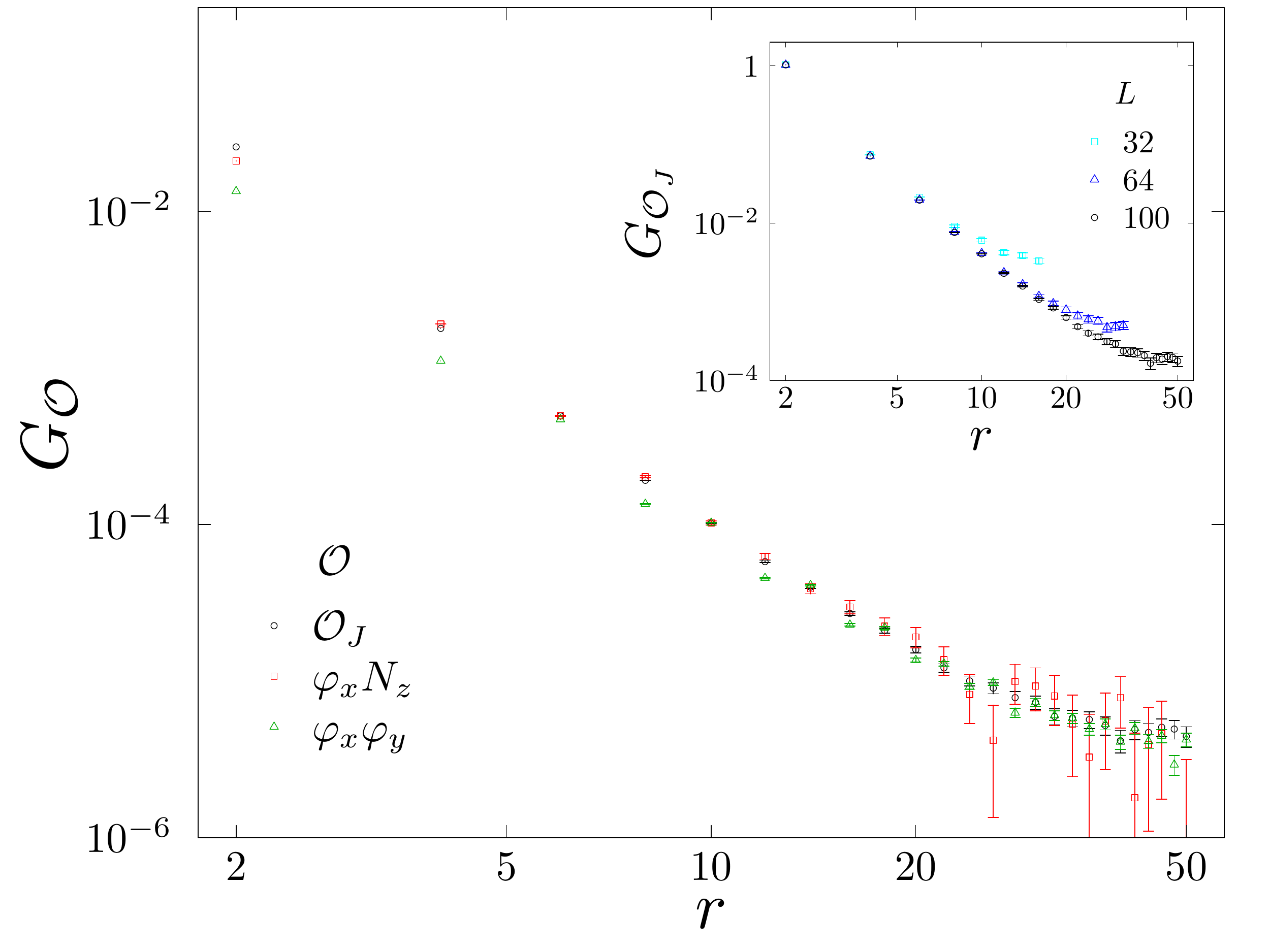}
 \end{center}
\caption{Correlators $G_\OO(r)$ for operators $\varphi_x N_z$,  $\varphi_x \varphi_y$ and $\OO_J$ in a system of size $L=100$. (The $G_\OO$ have been normalised to agree at $r=10$.) Inset: $G_{\OO_J}$ for various $L$.}
 \label{phi N correlation function}
\end{figure}

To test these predictions, Fig.~\ref{phi N correlation function} shows data for the two-point functions of $\OO_J$, $\varphi_x N_z$, and $\varphi_x \varphi_y$, or rather lattice versions of these operators. (See App.~\ref{lattice defns of operators} for definitions and Ref.~\cite{deconfined long paper} for a general discussion of correlation functions at the critical point.) Note the striking similarity of the three curves, as expected from $\mathrm{SO}(5)$. The slopes at $r\sim 10$ are around $x_2^\text{eff}\sim 1.5$, but this effective exponent may be strongly affected by finite size effects. (Various scaling dimensions, including the two-monopole dimension \cite{dyer et al}, are known at large $n$ in the $\mathrm{SU}(n)$ generalisation of Eq.~\ref{NCCP1 lagrangian} \cite{Monopole scaling dim, Kaul Sachdev large n, Monopole scaling dim 2, kaul sandvik large n}, and  show that a symmetry between N\'eel and VBS cannot persist in this limit.)

$\OO^{(4)}$ allows us to write both a subleading operator which breaks the symmetry between N\'eel and VBS ($\sum_{a=1}^3\sum_{b=4}^5 \OO^{(4)}_{aabb}$), and one which breaks the remaining symmetry for $\vec \varphi$  down to fourfold rotations ($\sum_{a=4}^5 \OO^{(4)}_{aaaa}$). Therefore it is possible that the same irrelevant exponent controls finite-size corrections to both types of symmetry enhancement (see also  App.~\ref{additional moment data}).

\textit{Nonlinear $\sigma$--model.} The  NL$\sigma$M description of the deconfined critical point proposed in Ref.~\cite{senthil fisher competing orders} is
\be\label{NLSM}
\mathcal{S}_\sigma = \int \dd^3 x \bigg( \f{1}{g} \big(\nabla \vec \Phi \big)^2 + \sum_i \lambda_i \mathcal{O}_i \bigg)
+ \mathcal{S}_\text{WZW}
\ee
where $\mathcal{S}_\text{WZW}$ is a topological Wess-Zumino-Witten term at level one (associated with the homotopy group $\pi_4(\mathrm{S^4}) = \mathbb{Z}$ of the target space). Physically, this term ensures that a vortex in the VBS order has an unpaired spin--1/2 at its core \cite{levin senthil, senthil fisher competing orders}. The $\mathcal{O}_i$ are the various anisotropies, some discussed above, that break $\mathrm{SO}(5)$ symmetry down to the microscopic physical symmetry. 

\begin{figure}[t]
 \begin{center}
 \includegraphics[width=0.8\linewidth]{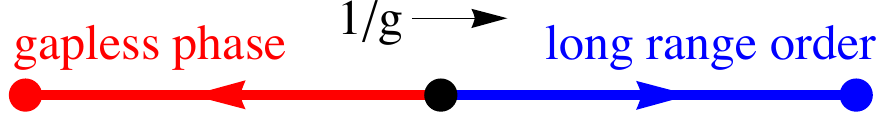}
 \end{center}
\caption{Conjectured phase diagram for NL$\sigma$M with WZW term in presence of full $\mathrm{SO}(5)$ symmetry. The fixed point on the left also governs the deconfined critical point, where $\mathrm{SO}(5)$ is emergent. Moving away from $J_c$ introduces the relevant symmetry-breaking perturbation $\OO_J$, leading to the N\'eel or VBS phase.}
 \label{phase diagram}
\end{figure}

Suppose that the N\'eel--VBS transition is continuous, with emergent $\mathrm{SO}(5)$ symmetry, and that the NL$\sigma$M is a viable description of this transition. Since the critical point is reached by tuning a single parameter, there is only a \textit{single} RG--relevant coupling in Eq.~\ref{NLSM}, namely the anistotropy $\OO_J$ of Eq.~\ref{relevant anistropy}. Therefore the $\mathrm{SO}(5)$---symmetric NL$\sigma$M, with no anisotropies, has a nontrivial infra-red \emph{stable} fixed point controlling a power-law correlated phase: see Fig.~\ref{phase diagram}. This fixed point  then also governs the deconfined transition. (The NL$\sigma$M also has a stable ordered phase even in the presence of the WZW term.)

The phase diagram of Fig.~\ref{phase diagram} is counterintuitive, as we are used to 3D $\sigma$--models with an ordered phase, a trivial disordered phase, and an unstable critical point in between. But that conventional picture applies to the NL$\sigma$M without a WZW term. The present results indicate that the WZW term causes the trivial disordered phase to be replaced with a stable critical phase. While the absence of a trivial disordered phase may seem surprising, it is in fact a necessary feature of any field theory which faithfully describes the spin-1/2 magnet, as argued in Ref.~\cite{senthil fisher competing orders}. By the 2D generalisation of the Lieb-Schultz-Mattis theorem \cite{hastings theorem, LSM}, the magnet cannot be in a trivially disordered phase with a unique, symmetric, gapped ground state (it can be critical, or topologically ordered, or spontaneously break a symmetry). Thus the field theory should not have such a trivial phase either \cite{anisotropies in field theory}. This argument does not by itself imply the existence of a stable critical phase, only the absence of a trivial fully disordered one. (See Refs.~\cite{senthil fisher competing orders, xu ludwig} for related discussions of the $\mathrm{SO}(4)$ case.)

The emergent $\mathrm{SO}(5)$ therefore indicates the existence of a 3D $\mathrm{SO}(5)$--symmetric CFT which is radically unlike standard Wilson-Fisher CFTs, in that there are no relevant singlet operators. It would be very interesting to investigate this using the conformal bootstrap \cite{O(n) bootstrap}, making use of numerical estimates for operator dimensions \cite{deconfined long paper}. This should be simpler \cite{multicritical footnote} than studying the critical $\nccp^1$ model without assuming $\mathrm{SO}(5)$.

\textit{Scaling violations.} The deconfined critical point shows strong violations of finite-size scaling. We argued in Ref.~\cite{deconfined long paper} that these are \emph{not} simply large corrections to scaling of the conventional type (i.e. from irrelevant or marginally irrelevant operators), but should instead be attributed \textit{either} to a first order transition with an anomalously large correlation length \textit{or} to a genuine critical point with unconventional finite-size scaling due to a dangerously irrelevant variable. The present results strongly suggest the second scenario --- a genuine continuous transition. This is because a critical point is the only natural explanation for the emergent $\mathrm{SO}(5)$ symmetry, which we have tested here to high precision. This symmetry therefore provides long-sought direct evidence that the deconfined transition is continuous.

In more detail: if one attempts to account for the data in terms of an anomalously weak first order transition (i.e. without postulating a genuine 3D critical point), one is led to a scenario where the {apparent} critical behaviour is due to a `nearby' fixed point at spacetime dimension slightly \textit{below} three \cite{deconfined long paper}. While this scenario can potentially explain  pseudo-critical behaviour up to an extremely large lengthscale, it cannot account for the emergent $\mathrm{SO}(5)$ symmetry --- this makes sense only for a 3D fixed point. While we can consider the $\nccp^{1}$ model in arbitrary dimension, the operator $\vec \varphi$ (interpreted for example as a monopole insertion operator) is special to 3D, and is required for construction of the $\mathrm{SO}(5)$ superspin. 

Assuming therefore that the transition is continuous, a possible explanation for the scaling violations is that a dangerously-irrelevant variable is required to cut off the fluctuations of a zero mode of the field \cite{deconfined long paper}. (This would be analogous to $\phi^4$ theory above four dimensions, where the quartic term is irrelevant, but cannot be set to zero since it is needed to prevent divergent fluctuations of $\phi$'s zero mode \cite{brezin zinn justin fss}.) This may suggest that an alternative field theory description exists which is more natural than the NL$\sigma$M \cite{AN TS to appear}.

\textit{Future directions.}  It would be interesting to investigate consequences of the emergent symmetry for finite temperature behaviour, as well as to look for signs of it using methods complimentary to Monte Carlo such as exact diagonalisation or DMRG.  The present results also motivate further analysis of $\mathrm{SO}(5)$--symmetric 3D CFTs --- for example with the conformal bootstrap --- and a more general investigation of the role played by WZW terms in field theories above two dimensions. For example, is there an analogous $\mathrm{SO}(6)$--symmetric CFT in 4D? Finally, we note that the critical behaviour at the deconfined transition remains perplexing, and deserves further investigation.

\textit{Acknowledgements.} A.N. thanks T. Senthil for very useful discussions and acknowledges the support of a fellowship from the Gordon and Betty Moore Foundation under the EPiQS initiative (grant no. GBMF4303). This work was supported in part by EPSRC Grant No. EP/I032487/1 and by Spanish MINECO and FEDER (UE) grant no. FIS2012-38206 and MECD FPU grant no. AP2009-0668.

\appendix

\section{Lattice definitions of operators}
\label{lattice defns of operators}

We refer to the lattice field theory of Ref.~\cite{deconfined long paper}, described there for $J=0$ but  modifiable for $J\neq0$. The N\'eel vector $N$ is defined on each link by $\vec N  =\zb^\dag \vec\sigma \zb$  with $|\zb|^2 = 2$ (these normalisations are purely a matter of convenience). A graphical expansion maps the lattice field theory to a partition function for loops taking two colours, red and blue \cite{cpn loops long}. Inserting operators on the links modifies the graphical expansion. Using $\Tr N_z z_c z_d^* =(2/3) \delta_{cd} ( \delta_{c1}-\delta_{c2})$, where $\Tr$ is the integral over  $\zb$s and $\Tr z_c z_d^* = \delta_{cd}$,  one may check that $N_z$ is simply the operator which measures the colour of a link in the loop gas: $N_z = (2/3) \chi$, where $\chi$ is $\pm 1$ depending on whether the link is red or blue. Therefore we obtain the probability distribution for the $z$-component of the uniform magnetisation, $N_z^\text{tot}$ by measuring the number of red links in the Monte Carlo simulation. We may measure $\vec \varphi$ simultaneously, allowing construction of the joint probability distribution. This is what we have done for system sizes $L\leq 100$. (In the main text the component we singled out was labelled $N_x$ rather than $N_z$, but of course by symmetry there is no difference.)

For larger system sizes we do not have data for the full probability distribution of link colours. However we have data for low moments of the loop length distribution, and this is sufficient to construct the moments $F_2^4$ and $F_4^4$ shown above.  We define, in a given configuration,
\be
S_k = \sum_\text{loops} (\text{length of loop})^k.
\ee
The necessary relations follow straightforwardly from $\sum_\text{links} N_z = (2/3)\sum_\text{links} \chi$. We express $\<(N_z^\text{tot})^s\>$ as a correlation function of $\chi$ operators in the loop gas. Summing over the possible colourings of a given loop configuration gives a nonzero result only if there are an even number of $\chi$ insertions on each loop. Taking into account the possible ways of assigning $\chi$s to loops, we have 
\ba
\< (N_z^\text{tot})^2\> & = (2/3)^2 \< S_2\>,\\
\< (N_z^\text{tot})^4\> & = (2/3)^4 \< 3S_2^2 - 2 S_4\>.
\end{align}
Importantly, these relations remain true if powers of $\vec\varphi$ are inserted on the left and right hand sides. This allows us to construct the abovementioned moments. 

Note that in order to give optimal statistics, the uniform magnetisation/VBS order is always defined by an integral over the three-dimensional sample --- i.e. space-time, in the quantum language --- rather than a two-dimensional `spatial' slice. 

The 3D `L' lattice \cite{cardy network models review} on which the model is defined is bipartite, with $\varphi_x$ defined at nodes of the A sublattice and $\varphi_y$ at nodes of the B sublattice, and the Boltzmann weight couples nearest neighbour $\varphi$ components on the same sublattice \cite{deconfined long paper}. The operators whose correlation functions are plotted in Fig.~\ref{phi N correlation function} are all defined at a node, and the correlation functions are evaluated for pairs of nodes on the same sublattice, separated parallel to a coordinate direction. We define $\OO_J(r)$ as the sum of the eight terms in the lattice energy which involve the node $r$, and subtract the average so that $\< \OO_J \> = 0$. The operator $[\varphi_x \varphi_y](r)$ is defined as the sum of $\varphi_x(r) \varphi(r')$ over the four nearest neighbours $r'$ of $r$ (see Fig.~1 in Ref.~\cite{deconfined long paper}) and $[\varphi_x N_z](r)$ is defined as the sum of $\varphi_x(r)N_z(l)$ over the four links $l$ connected to $r$, with $N_z$ defined as above in terms of link colours. Each of these  operators is therefore supported on star: we take the separation of the two points in the correlator to be perpendicular to the plane of the stars.

\begin{table}[t]
\begin{center}
\begin{tabular}{|r|r|r|r|r|}
\hline
$F_l^a$ & $a = 0$ & $a=2$ & $a=4$ & $a=6$\\
\hline
$l=2$ &0.00035(17)& 0.000(4) & 0.00(2) & 0.01(7)\\
$l=4$ &0.0002(7) & -0.001(2) & -0.004(8) & -0.02(3)\\
$l=6$ & 0.0002(5)& 0.000(1) & -0.003(5) & -0.02(2)\\
\hline
\end{tabular}
\end{center}
\caption{Moments $F_l^a$ in the $(\widetilde N_x, \widetilde\varphi_x)$ plane for $L=100$ (taken at $J=0.0885$).}
\label{Nxpx plane}
\end{table}

\begin{figure}[t]
 \begin{center}
 \includegraphics[width=0.9\linewidth]{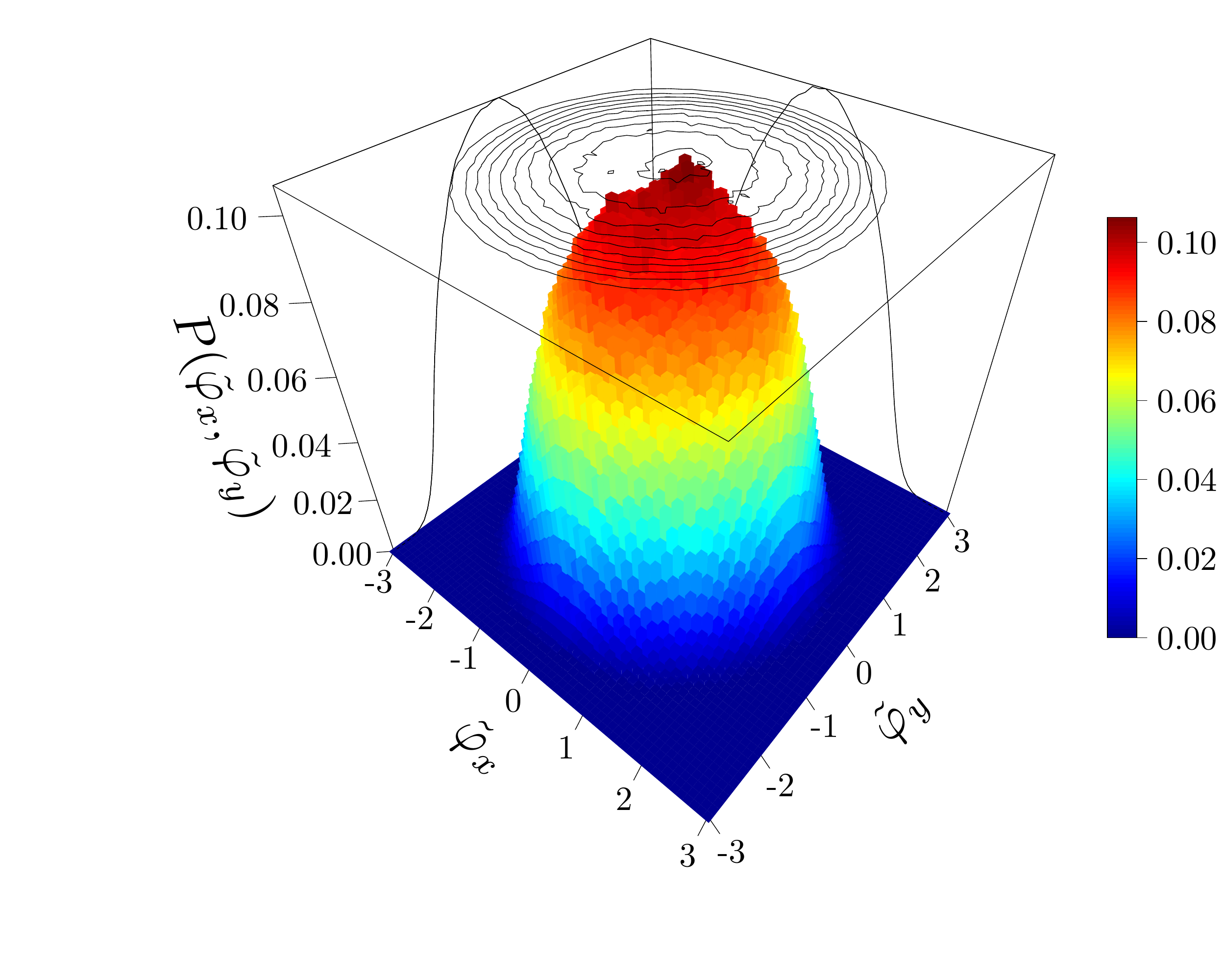}
 \end{center}
\caption{The joint probability distribution $P(\widetilde\varphi_x, \widetilde\varphi_x)$, after rescaling $\varphi_x$ and $\varphi_y$ to have unit variance, in a critical system of size $L=100$. Upper plane shows contour plot.}
 \label{probability distribution in phi plane}
\end{figure}

\section{Additional data for probability distribution}
\label{additional moment data}

Table~\ref{Nxpx plane} shows data for a variety of the moments $F_l^a$ defined in Eq.~\ref{moments eq} in a system of size $L=100$ at $J=0.0885$. These quantify the $\mathrm{U}(1)$ symmetry of the distribution in the $(\nx, \fx)$ plane. All are consistent with zero.

\begin{table}[h]
\begin{center}
\begin{tabular}{|r|r|r|r|r|}
\hline
$L$ & $a=0$ & $a=2$ & $a=4$ & $a=6$ \\
\hline
32 & -0.0086(9) &-0.035(3) & -0.166(10) & -0.88(5) \\Ê
64 &  -0.005(1) &-0.021(3) & -0.095(11) & -0.48(5) \\Ê
100 & -0.0028(14)& -0.014(4) & -0.069(15) & -0.35(7)\\
\hline
\end{tabular}
\end{center}
\caption{Moments $\widetilde F_4^a$ in the $(\widetilde\varphi_x,\widetilde\varphi_y)$ plane, quantifying fourfold anisotropy for the VBS, for $L=32,64,100$. (Note that  moments grow with $a$ for trivial reasons.)}
\label{varphi plane L}
\end{table}%

We also examine the symmetry properties of the probability distribution in the  $(\fx, \widetilde\varphi_y)$ plane. This is shown in Fig.~\ref{probability distribution in phi plane} for a critical system of size $100$. Again it is $\mathrm{U}(1)$ symmetric to an excellent approximation. Interestingly though, the finite-size corrections to this $\mathrm{U}(1)$ are larger than those for the $\mathrm{U}(1)$ in the $(\nx, \fx)$ plane (but still small). Table~\ref{varphi plane L} quantifies fourfold anisotropy for the VBS at the critical point via the moments $\widetilde F_4^a$ (defined analogously to $F_l^a$ but in the $(\widetilde\varphi_x, \widetilde\varphi_y)$ plane). They decrease with system size roughly as $L^{-c}$ with $c\sim 0.8$.

In Fig.~\ref{moments as a function of L} of the main text the straight lines indicate the expected values of the moments $F_2^4$ and $F_4^4$ within the phases. These values follow straightforwardly. The vector which is in disordered phase has a Gaussian distribution, so its moments follow from Wick's theorem. The other order parameter can be treated as fixed in magnitude and averaged over symmetry-equivalent directions. In the N\'eel phase this gives $F_2^4=F_4^4=-1.2$. Our choice of coordinates is such that deep in the VBS phase $\vec \varphi \propto (\pm 1, \pm 1)$, which gives $F_2^4=2$, $F_4^4=-2$. For weak VBS order there is a large range of $L$ where $\mathrm{U}(1)$ symmetry for $\vec \varphi$ survives, and $F_2^4 =1.5$, $F_4^4=-1.5$.


\begin{thebibliography}{200}
\bibitem{eta footnote} Early numerical work suggested that $\eta_\text{N\'eel} = \eta_\text{VBS}$  \cite{Sandvik JQ}, but this was subsequently questioned \cite{lou sandvik kawashima}.
\bibitem{scaling dim footnote} At a conventional critical point, $L$-independence of $\sigma_\varphi/\sigma_N$ would be equivalent to equality of $\eta_\text{VBS}$ and $\eta_\text{N\'eel}$, but finite-size scaling may be unconventional here \cite{deconfined long paper}.
\bibitem{anisotropies in field theory} The anisotropies are not important for this argument since if a trivial disordered phase existed in their absence it should be stable to their inclusion \cite{senthil fisher competing orders}.
\bibitem{multicritical footnote} If only  $\mathrm{SO}(3)$ symmetry is assumed the theory must be treated as a high-order multicritical point, as in addition to the relevant mass term there are relevant $\mathrm{SO}(3)$-invariant operators for single/double monopoles. (Lattice symmetry forbids them at the N\'eel-VBS transition.)
\end{thebibliography}
\end{document}